\documentclass{article}
\usepackage[utf8]{inputenc}
\usepackage{xcolor}
\usepackage{float}
\usepackage[justification=centering]{caption}
\usepackage[margin=0.65in]{geometry}
\usepackage[pdftex]{graphicx}
\graphicspath{{\Documents}}     
\DeclareGraphicsExtensions{.pdf,.jpeg,.png}
\usepackage{subfigure}
\usepackage{titlesec}
\usepackage{amssymb,amsmath}
\usepackage{moreverb}
\usepackage{multirow}
\usepackage[utf8]{inputenc}
\usepackage{multicol}
\usepackage{array}
\usepackage{url}

\title{\textbf{An ensemble of convolution-based methods for fault detection using vibration signals}}

\author{Xian Yeow Lee\thanks{denotes equal contribution.}
, Aman Kumar*, Lasitha Vidyaratne*, Aniruddha Rajendra Rao*,\\ Ahmed Farahat, Chetan Gupta\\
Industrial AI Lab, Hitachi America Ltd., R\&D}

\begin{document}
\maketitle
\vspace{-0.5cm}
\hspace{0.5 cm}\{xian.lee, aman.kumar, lasitha.vidyaratne, aniruddha.rao, ahmed.farahat, chetan.gupta\}@hal.hitachi.com\\

\hspace{0.35cm}\textbf{Keywords:} Time Series, Classification, Convolution, Ensemble, Vibration Data.

\abstract{This paper focuses on solving a fault detection problem using multivariate time series of vibration signals collected from planetary gearboxes in a test rig. Various traditional machine learning and deep learning methods have been proposed for multivariate time-series classification, including distance-based, functional data-oriented, feature-driven, and convolution kernel-based methods. Recent studies have shown using convolution kernel-based methods like ROCKET, and 1D convolutional neural networks with ResNet and FCN, have robust performance for multivariate time-series data classification. We propose an ensemble of three convolution kernel-based methods and show its efficacy on this fault detection problem by outperforming other approaches and achieving an accuracy of more than 98.8\%.
}

\section{Introduction}

Prognostics and health management (PHM) is concerned with the prediction of a potential failure or health degradation of  equipment and the management of actions needed to alleviate such conditions. PHM involves solving several tasks such as fault detection and isolation, failure prediction and remaining useful life estimation (RUL)~\cite{intro1, HU2022108063}. PHM has been a crucial field with a variety of applications in several industries. Several approaches have been proposed to perform PHM, which can be categorized into model-based approaches and data-driven approaches. The former use physics-based models to represent the degradation of the equipment and the latter uses historical data of system measurements to learn degradation patterns. Examples of data-driven approaches include traditional machine learning and, recently, deep learning models. 

 
 In this paper, we focus on solving a fault detection problem using vibration data collected from planetary gearboxes in a test rig as part of the ICPHM 2023 Data Challenge~\cite{icphm}. The vibration data consists of multivariate time series data of accelerometer measurements in the $x$, $y$, $z$ directions, and the problem can be formulated as a multivariate time-series classification problem. Over the past few years, several approaches have been proposed for multivariate time-series classification, with many state-of-the-art techniques emerging as a result. This includes traditional machine learning based approaches such as distance based, and feature-driven methods, and deep learning methods that perform feature learning from data. Distance based methods involve dynamic time warping techniques~\cite{muller2007dynamic, shokoohi2017generalizing, mei2015learning} that essentially perform stretching and compressing operations in the time axis to compute optimal alignment between time-series. Feature-driven approaches utilize a hand-engineered feature extraction scheme to compute a reduced set of static features from time-series data. These features can then be used with traditional classification methods. The standard features extraction methods can range from simple summary statistics based~\cite{barandas2020tsfel, fulcher2017feature, christ2018time}, to functional data analysis based techniques~\cite{10020482, fda, fdabook}, to wavelet and shapelet based techniques~\cite{chaovalit2011discrete, rhif2019wavelet, lines2012shapelet, bagnall2020usage, ye2011time}. Of these techniques, recently developed methods such as Hierarchical Vote Collective of Transformation-based Ensembles (HIVE-COTE)~\cite{bagnall2020usage} and the Random Convolutional Kernel Transform (ROCKET)~\cite{dempster2020rocket} have shown promise in a variety of multivariate time-series classification applications. In contrast to these traditional approaches, deep learning based methods offer the ability to learn the classification task from time-series data directly. Recurrent neural networks~\cite{schuster1997bidirectional} and its variants such as long short-term memory (LSTM)~\cite{hochreiter1997long}, and gated recurrent units (GRU)~\cite{cho2014learning} are specifically designed to process time-series through temporal recurrence.
 
Convolution kernels for time series data are filters that are applied to the input data in a sliding window fashion. These filters are typically small, fixed-sized arrays of values that are multiplied element-wise with the corresponding input values and then summed to produce a single output value. The convolution operation can be used to extract different types of features from the time series data~\cite{dempster2020rocket}, such as local patterns, trends, and periodicity. Different types of convolution kernels can be used depending on the type of features that are desired. The success of ROCKET and its variants~\cite{minirocket, Tan2021MultiRocketES} is largely due to the use of convolution kernels to extract features. Recent studies have proposed the adaptation of convolutional neural networks with a 1-dimensional convolution kernel for time-series data processing due to the significant compute efficiency associated with convolutional neural architectures. Of these, 1D residual network (ResNet) type architectures and 1D fully convolutional network (FCN) type architectures have shown robust performance for multivariate time-series data classification~\cite{wang2017time}.
 
After conducting preliminary data analysis, we identified several state-of-the-art techniques and applied them to the initial dataset released for model development. We then identified the three top-performing models, all of which shared a common theme of using convolutional kernels, and developed an ensemble technique to combine the results of these three models.

\section{Data Description}\label{sec:data_description}
This section summarizes the test rig set-up used to generate the dataset studied in this work. The test rig comprises a driving motor, a two-stage planetary gearbox, a two-stage parallel gearbox, and a magnetic brake. The data focuses on the planetary gearbox. The motor, which is driven by an alternating current, controls the test rig's speed ranging from 0 to 6000 revolutions per minute (rpm) via a speed transducer. The magnetic brake applies torque to the output shaft of the parallel gearboxes. In practical applications, the sun gear teeth on the second stage of the planetary gearbox are prone to damage due to heavy alternative stress and impact, thus creating four common sun gear faults - surface wear, chipped, crack, and tooth missing. Vibration signals were collected using the National Instruments portable data acquisition system, and the acceleration sensors mounted on the second stage of the planetary gearbox was manufactured by PCB Piezotronics. The vibration signals were obtained under two different operating conditions, one with a motor speed of 1500 rpm and 10 Nm load, the other setting being 2700 rpm motor speed and 25 Nm load. In this paper, we refer to two datasets as operating condition 1500 rpm and operating condition 2700 rpm. Each vibration signal was recorded in the $x$, $y$, and $z$ directions for five minutes with a sampling frequency of 10 kHz. The vibration signals have been divided into smaller segments using a window size of 200 data points.

The dataset provided for this challenge contains 50000 vibration signals, their corresponding ground-truth labels, and two different operational conditions. The signals include both normal (Class 0) and four different fault conditions (Crack - Class 1, Surface Wear - Class 2, Chipped - Class 3, Tooth Missing - Class 4). Each class in the data has 10000 samples. The given data has no missing value and is well-balanced.

\section{Exploratory Data Analysis}\label{sec:eda}

\begin{figure}[h]
    \centering
    \includegraphics[width=0.89\columnwidth]{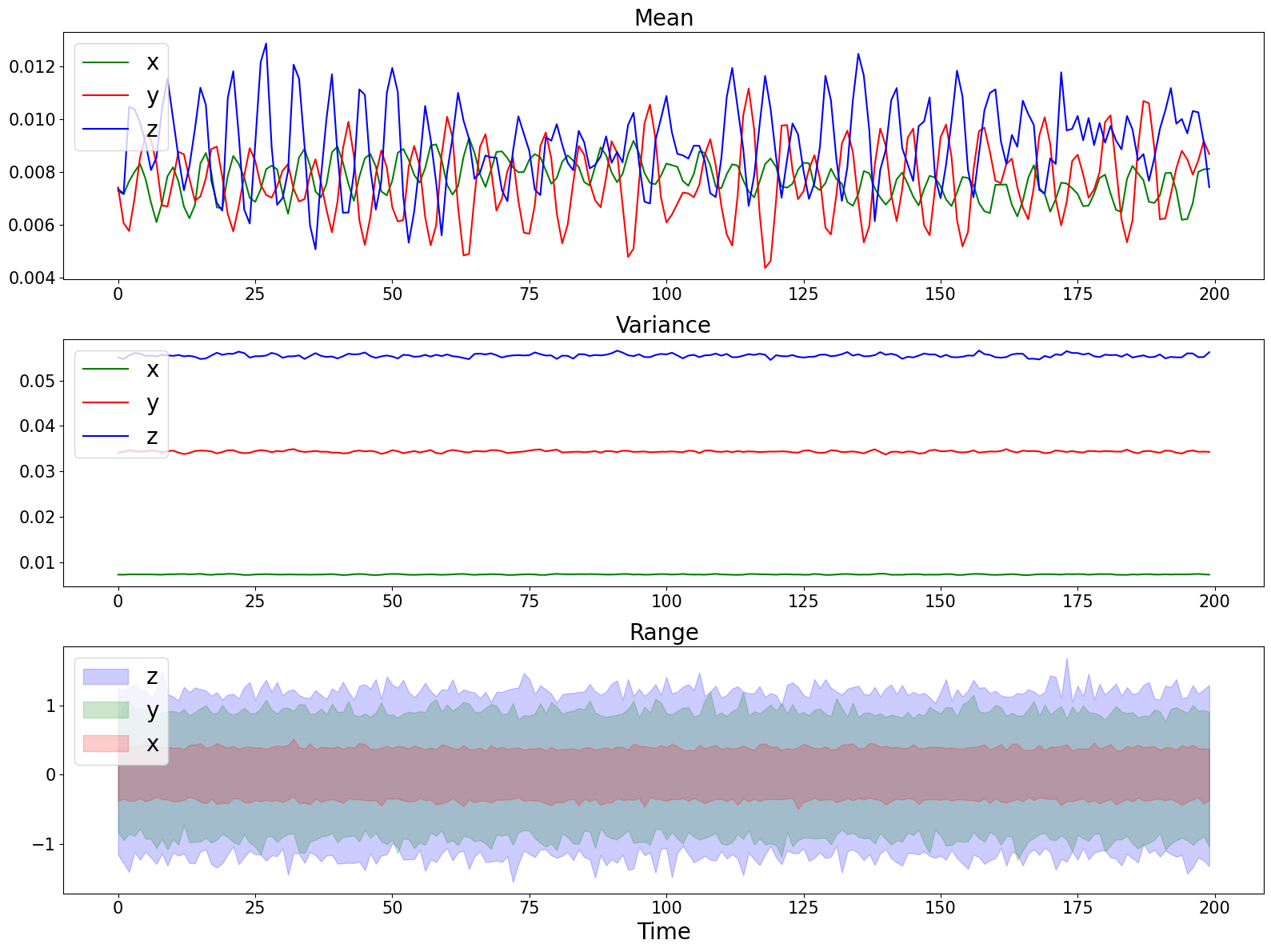}
    \caption{Mean, variance and range of the vibration data for operating condition 1500 rpm. }
    \label{fig:stats_15}
\end{figure}

\begin{figure}[h]
    \centering
    \includegraphics[width=0.89\columnwidth]{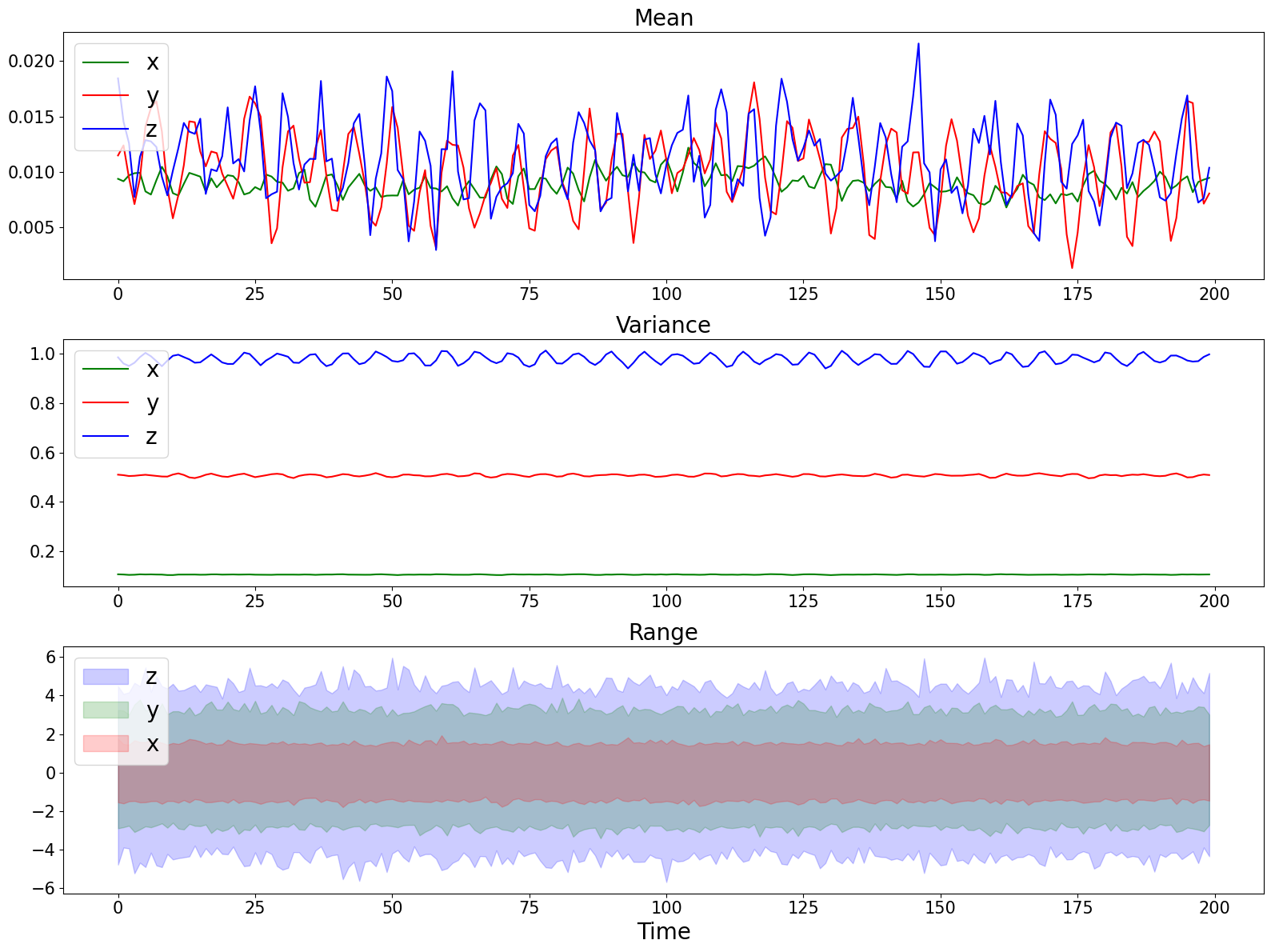}
    \caption{Mean, variance and range of the vibration data for operating condition 2700 rpm. }
    \label{fig:stats_27}
\end{figure}

In this section, we provide a simple exploratory data analysis and a brief discussion which leads to the choice of modeling methods in the next section. Since the data we are studying in this work had an equal number of samples per class and were all collected with the same frequency, as described in Section~\ref{sec:data_description}, we did not have to consider factors such as class or distributional imbalance. In Figures~\ref{fig:stats_15} and~\ref{fig:stats_27}, we plot the means, variances, and ranges of each channel for the data in the operating conditions of 1500 rpm and 2700 rpm, respectively. Based on the figures, we observed that the means and variances in each channel for both operating conditions are largely constant with no observable trends, signifying that the time series data are stationary. Furthermore, we can also observe that data collected in the $z$-channels had the largest range, followed by the data in the $y$-channels and, lastly, the $x$-channels for both operating conditions, though they remain largely within the same order of magnitude. 

\begin{figure}[h]
    \centering
    \includegraphics[width=0.89\columnwidth]{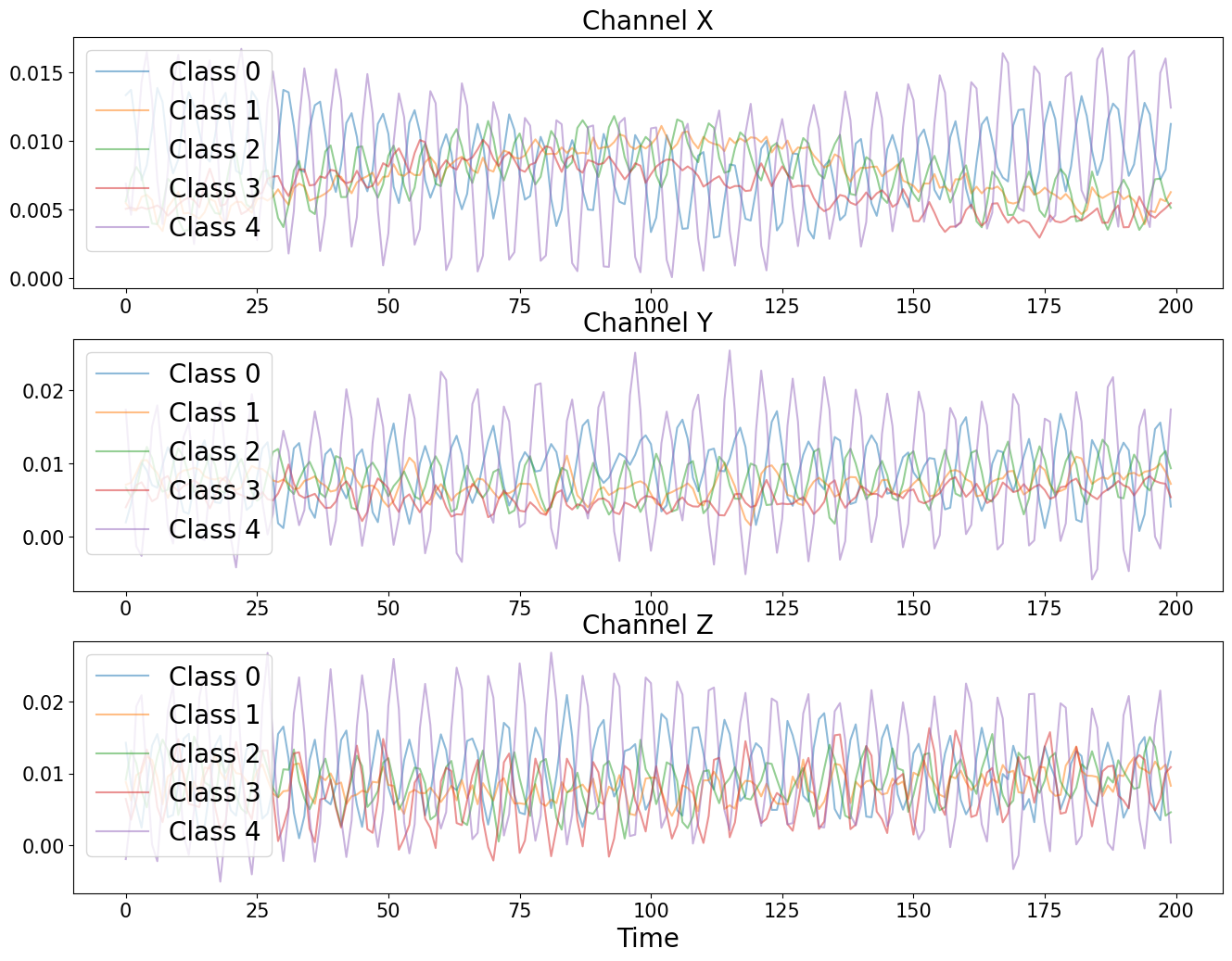}
    \caption{Class-specific mean values of each channel of the vibration data for operating condition 1500 rpm. }
    \label{fig:class_15}
\end{figure}

\begin{figure}[h]
    \centering
    \includegraphics[width=0.89\columnwidth]{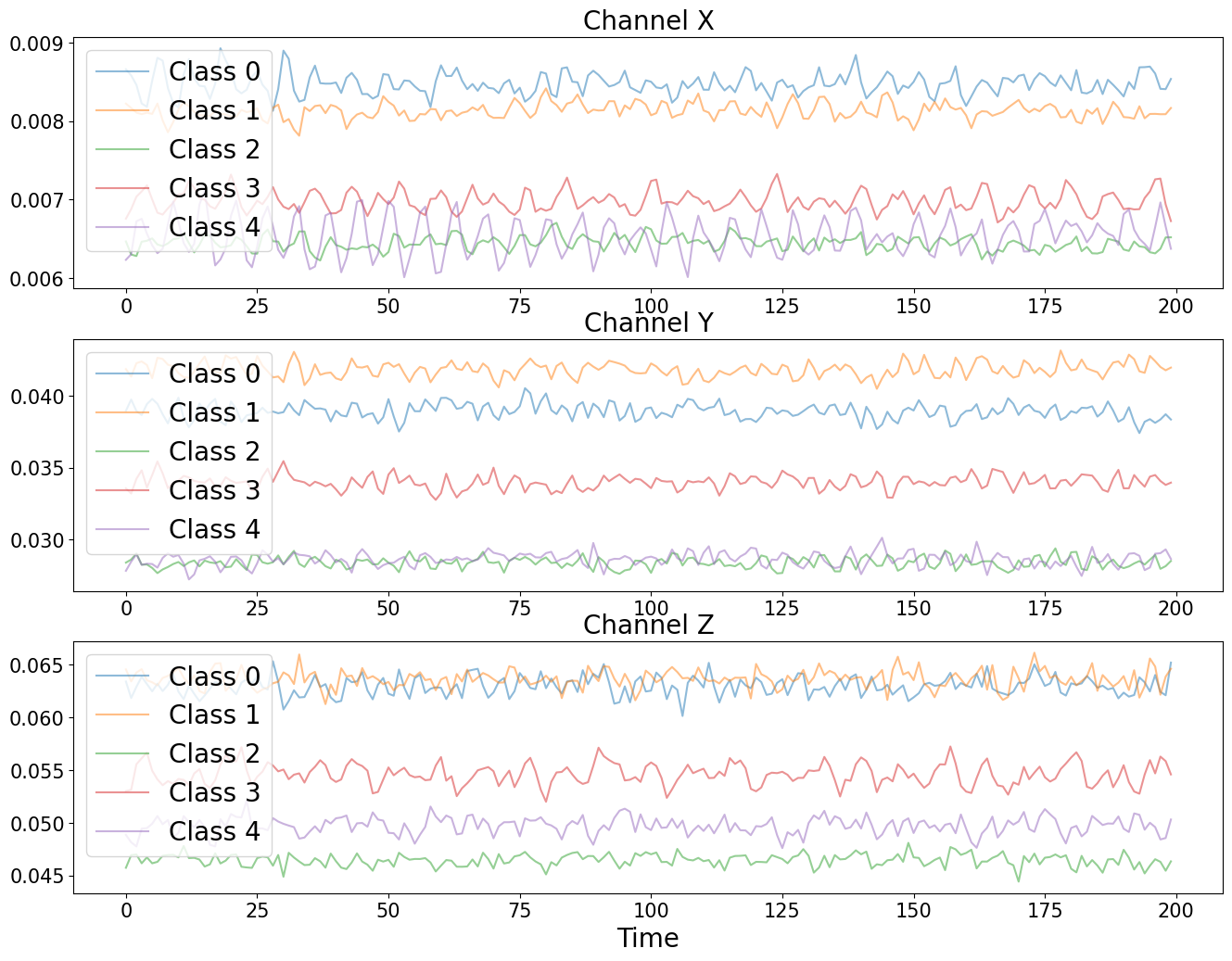}
    \caption{Class-specific variance values of each channel of the vibration data for operating condition 1500 rpm. }
    \label{fig:var_15}
\end{figure}

\begin{figure}[h]
    \centering
    \includegraphics[width=0.89\columnwidth]{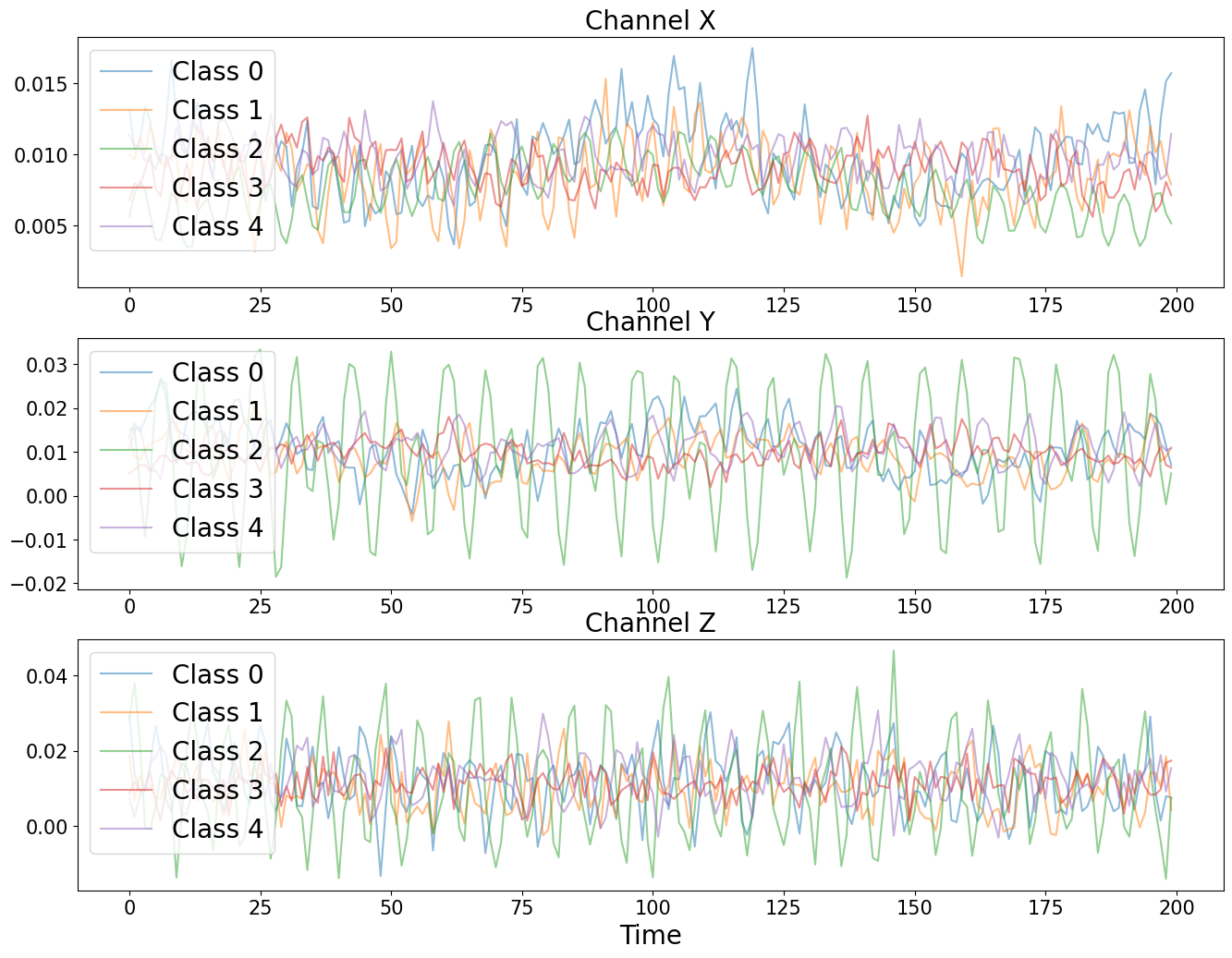}
    \caption{Class-specific mean values of each channel of the vibration data for operating condition 2700 rpm.}
    \label{fig:class_27}
\end{figure}

\begin{figure}[h]
    \centering
    \includegraphics[width=0.89\columnwidth]{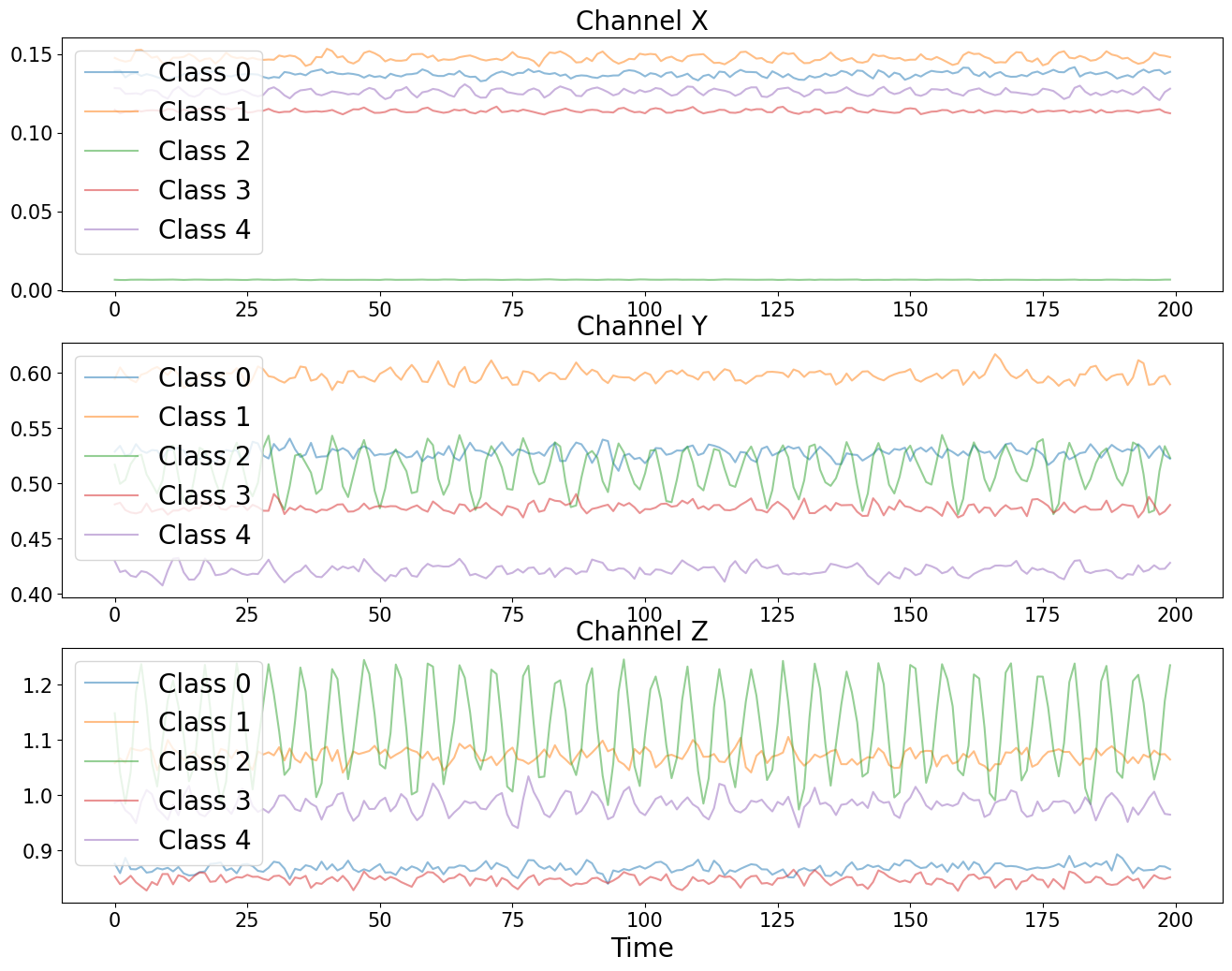}
    \caption{Class-specific variance values of each channel of the vibration data for operating condition 2700 rpm. }
    \label{fig:var_27}
\end{figure}

Next, we plot the means of data of each class and visualize them in Figures~\ref{fig:class_15} and~\ref{fig:class_27}. In both operating conditions, we can observe that the means of the vibration for each class and every channel are highly overlapping. Lastly, Figures~~\ref{fig:var_15} and~\ref{fig:var_27} illustrate the variances of the data for operating conditions at 1500 rpm and 2700 rpm, respectively. Interestingly, in the operating condition of 1500 rpm, we can observe that the vibrations of each class of defect mostly exhibit a different magnitude of variance. For example, in Figure~\ref{fig:var_15}, we can clearly see that vibrations associated with class 0 exhibit the most variance, followed by class 1, 3, 4, and 2 in channel $x$. A mixed observation can be made for the vibration data collected in the other channels for the operation condition of 2700 rpm, as seen in Figure~\ref{fig:var_27}, where the variances are clearly separated by the classes, though in a different order. Nonetheless, since the mean values are highly overlapping, we conclude that a simple linear classifier or a naive distance-based algorithm such as the k-nearest neighbor algorithm operating on the raw features of the data wouldn't yield a competitive classification performance. Hence, in the following section, we explore approaches that leverage automated feature extraction or feature learning for better performance. 

\section{Modeling}

In this section, we discuss the several known approaches that have been reported to yield state-of-the-art performance for time-series classification, specifically 1D and 2D convolution-based methods, LSTM-based methods, automated feature extraction combined with traditional machine learning classifiers such as random forest, logistic regression, and k-nearest neighbors as well as hybrid approaches that combine one or more of these methods. Note that most of the methods above will readily accept the raw time series data as input and do not require any pre-processing to prepare the data to be ingested by the models, with the exception of the 2D-convolution methods. To pre-process the time-series data for 2D-convolution, we performed a short-time Fourier transform on the time series to obtain a spectrogram that contains frequency and time information. We then use the spectrogram as image inputs for the 2D-CNN model. In addition, we also experimented with simple data augmentation techniques such as adding bounded noise to the time series data to increase the number of training data points. However, in the context of this vibration dataset, we find that the number of training data is sufficiently large, and using data augmentation did not yield any additional improvements in terms of accuracy. 

Overall, we find that while most methods generally resulted in decent classification performance with cross-validation accuracy greater than 90\%, three particular methods consistently outperform the rest. We discuss these three methods in detail below. 

\subsection{Multi-scale 1D ResNet}

This model architecture is based on a simpler variation of CSI-Net~\cite{wang2018csi} that was developed as a lighter-weight version for time-series classification. The main idea of this method is to extract features of multiple scales along the time axis from the time series data using 1D-convolution kernels of different sizes and combining the multi-scale features to make a classification. 

Specifically, the model architecture consists of an initial shared convolution block that is constructed from a 1D-convolution, batch normalization, and max-pooling layer. The output of this convolution block is then processed by three parallel feature extractors. These feature extractors consist of three blocks of 1D-convolution, batch normalization, and ReLU activation functions with residual connections and average pooling after the final block to generate a 256-dimensional feature vector. The only difference between each of these feature extractors is that they used a different kernel size for the 1D-convolution layer, specifically a kernel size of 3, 5, and 7. The extracted features are then concatenated together to form a 768-dimensional feature vector before being processed by a fully-connected and softmax layer to make a classification. 

To improve the generalizability of the implementation, we also included a triplet margin loss~\cite{balntas2016learning,chechik2010large} to the standard cross entropy loss function and used a learning rate scheduler during training to decay the learning rate. We empirically find that including these two modifications improved the overall validation and testing accuracy of this method. 

To train this model, we up-sampled the vibration data in all three channels from 200 points to 512 points using a linear interpolation to fit the input dimensionality of the model. Furthermore, we also used Adam optimizer with an initial learning rate of 0.001 and a learning rate scheduler which reduces the learning rate by a factor of 0.1 when the validation accuracy plateaus. We trained the model for 100 epochs and saved the model with the best validation accuracy. For additional details, we refer the readers to the implementation that is based on~\cite{ResNetCode}.

\begin{figure*}[h]
    \centering
    \includegraphics[width=0.8\linewidth]{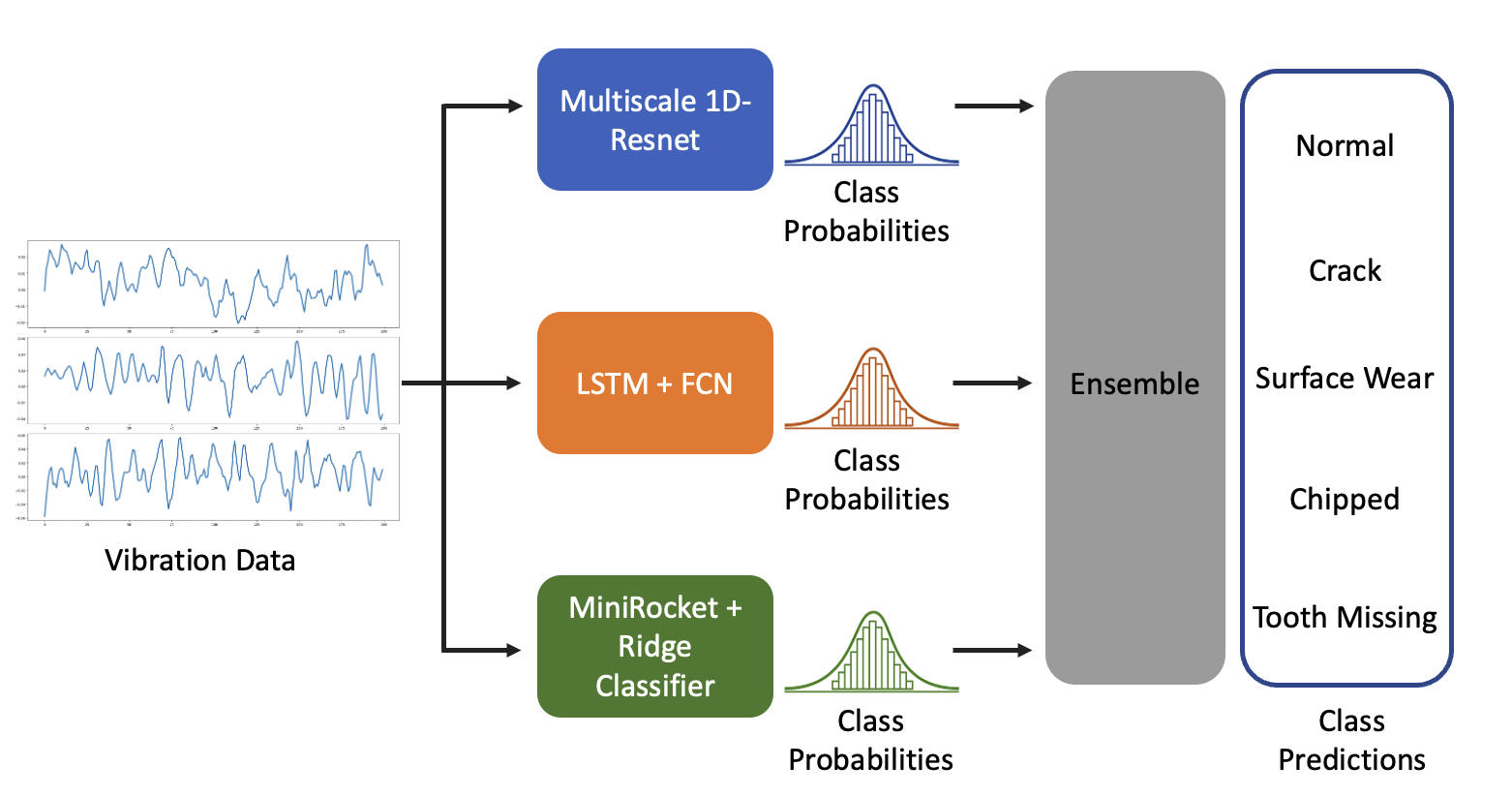}
    \caption{Overview of our approach, which combines two deep learning methods and an automated feature extraction method with a ridge classifier via an ensemble strategy to make a final defect classification.}
    \label{fig:overview}
\end{figure*}

\subsection{LSTM-FCN}
Long Short-Term Memory (LSTM) networks are a traditional class of neural networks that has proven to improve upon recurrent neural networks (RNN) due to its capability of addressing the vanishing gradient problem through gating functions~\cite{hochreiter1997long}. An LSTM maintains a memory vector and a hidden vector responsible for controlling outputs and updates of the state at each time step and can learn temporal dependencies in sequences~\cite{graves2012supervised}.

For problems pertaining to time series multi-class classification, the use of temporal convolutions or 1D-convolutions have shown to be an effective learning model~\cite{wang2017time}. The Fully Convolutional Network (FCN) is comprised of temporal convolutions and global average pooling, which are used for feature extraction and reduction of parameters prior to classification, respectively~\cite{lin2013network}. 

We have used an architecture consisting of a fully convolutional block (present in CNN architecture), which is augmented by an LSTM block and dropout layers. The fully convolution block contains the stacking of three temporal convolutional blocks (1D-convolution) having filter sizes 128, 256, and 128, where every block is accompanied by batch normalization and a ReLU activation function. The final convolution block is followed by global average pooling.

In parallel, the time series input is sent to a dimension shuffle layer, which transforms the input by converting any univariate time series into a multivariate time series with a single step, prior to passing through the LSTM block and a dropout layer. This output, along with the output from the previously obtained global average pooling layer, is concatenated and passed through a softmax layer for the final classification. The loss function for classification is the negative log-likelihood or cross-entropy loss.

Similar to the previously discussed multi-scale ResNet, we have used Adam optimizer, a learning rate scheduler, and trained the model for 100 epochs. We found that the learning rate scheduler has helped to stabilize the training and improved overall generalizability, specially in the training of operating condition 1500 rpm data where we observed a substantial variance. The implementation of LSTM-FCN is based on~\cite{LSTMFCN}.

\subsection{ROCKET}
ROCKET (RandOm Convolutional KErnel Transform)~\cite{dempster2020rocket} is a fast and effective method for time series classification. It is based on the idea of applying random convolutional filters to the time series data and then feeding the resulting features into a linear classifier. This approach is based on the insight that linear models can be very effective for high-dimensional data when combined with appropriate feature engineering.

The random convolutional filters used in Rocket are essentially random projections of the time series data onto a small set of fixed random vectors. The resulting features are then processed using the max-pooling operation, which identifies the maximum value within each feature map, and concatenates them into a single feature vector.

Rocket has been shown to be competitive with state-of-the-art time series classification methods while being significantly faster and more scalable. It also has the advantage of being simple to implement and easy to interpret since the resulting features can be visualized and analyzed to gain insight into the characteristics of the time series data.

MiniRocket is a configured version of Rocket~\cite{dempster2020rocket} that is almost deterministic, offering comparable accuracy while being 75 times quicker on larger datasets. MiniRocket~\cite{minirocket} is a fast and accurate feature extraction technique for time series classification. MiniRocket has been shown to outperform state-of-the-art methods on several benchmark time series datasets while also being much faster than traditional feature extraction techniques like Fourier transforms and wavelets. Additionally, MiniRocket can be easily parallelized and scaled to large datasets.

In comparison to Rocket, MiniRocket uses a fixed set of convolutional kernels with fewer hyper-parameters. This reduces the number of kernel options and maximizes computational efficiency while maintaining accuracy. Unlike Rocket, MiniRocket only uses kernels of length 9 and restricts the values of the weights to either -1 or 2, with a small, fixed set of 84 kernels that are almost entirely deterministic. MiniRocket also limits the maximum number of dilations per kernel to 32 to maintain efficiency. The only random component of MiniRocket is the bias value, which is drawn from the quantiles of the convolutional output from one random training example.

We implement MiniRocket~\cite{MiniRocketcode} using the module available on sktime~\cite{markus_loning_2022_7117735}. It is used to generate approximately 10000 features, and then we use Ridge Classifier to perform the multi-class classification.

\begin{table*}[!h]
\renewcommand{\arraystretch}{1.3}

    \centering
    \begin{tabular}{l|c c c c c |c c}
    \hline
    \hline
    Method  & Fold 1 & Fold 2  & Fold 3 & Fold 4 & Fold 5 & Average Accuracy & Average Training Time (s) \\
    \hline
    Multi-scale 1D ResNet  &  98.50 &	98.56 &	98.59 &	98.44 &	98.51 & 98.520 $\pm$ 0.058 & 799\\
    LSTM-FCN   & 98.03 & 97.24 & 97.05 & 96.36 & 97.74 & 97.284 $\pm$ 0.579 & 524\\
    MiniRocket  & 97.95 & 97.54 & 97.61 & 97.89 & 97.48 & 97.694 $\pm$  0.212 & 542\\
    \textbf{Ensemble (Average)} & 98.97  & 98.78 & 98.84 & 98.74 & 98.99 & \textbf{98.864 $\pm$ 0.112} & -\\
    Ensemble (Max) & 98.88 & 98.72 & 98.88 & 98.73 & 99.05 & 98.852 $\pm$ 0.135 & - \\
    \hline
    \hline
    \end{tabular}
    \caption{Five-fold cross-validation results for operating condition 1500 rpm. Method with best overall accuracy highlighted in bold.}
    \label{tab:1500_results}
\end{table*}

\begin{table*}[!h]
\renewcommand{\arraystretch}{1.3}
    \centering
    \begin{tabular}{l|c c c c c |c c}
    \hline
    \hline
    Method  & Fold 1 & Fold 2  & Fold 3 & Fold 4 & Fold 5 & Average Accuracy & Average Training Time (s)\\
    \hline
    Multi-scale 1D ResNet  &  98.39 &	98.41 & 98.44 & 98.43	& 98.39 & 98.412 $\pm$ 0.023 & 796\\
    LSTM-FCN   & 99.05 & 98.88 & 98.80 & 98.82 & 98.94 & 98.890 $\pm$ 0.090 & 522\\
    MiniRocket  & 97.31 & 97.59 & 97.52 & 97.50 & 97.55 & 97.494 $\pm$ 0.108 & 559\\
    \textbf{Ensemble (Average)} & 99.01  & 98.86 & 98.84 & 98.74 & 98.99 & \textbf{98.968 $\pm$ 0.064} & - \\
    Ensemble (Max) & 98.91 & 98.86 & 98.95 & 99.02 & 99.01 & 98.950 $\pm$ 0.067 & - \\
    \hline
    \hline
    \end{tabular}
    \caption{Five-fold cross-validation results for operating condition 2700 rpm. Method with best overall accuracy highlighted in bold.}
    \label{tab:2700_results}
\end{table*}

\subsection{Ensemble}
Ensemble approaches are commonly used to improve classification results. Two common methods are to take either the average or maximum probability value for classification. By combining the output from multiple models, we can make use of the different aspects being learned in each one. In our case, we are working with three different models and want to utilize their unique strengths through the ensemble. Ultimately, this approach can increase accuracy and provide more robust predictions. 

Our overall approach is summarized in Figure~\ref{fig:overview}. We feed the vibration data from all three channels into the three individual models described above and perform the required pre-processing and model training accordingly. Rather than using a single model's output to perform a classification, we take the predicted class probabilities from each model and combine them together to make a final classification, either by directly taking the maximum class probability or averaging the class probabilities and taking the maximum average probability. 

\section{Results}

\subsection{Accuracy comparison}
We conducted a five-fold cross-validation experiment on the vibration data to assess the performance of both individual methods and our proposed ensembling approach. For each fold, we reported the accuracy, the overall average accuracy and standard deviation for each approach. To split the data into the folds, we performed a 70/10/20 split, corresponding to the training, validation, and testing data.

In Table~\ref{tab:1500_results}, we tabulate the test scores of our experiments for the data with operating condition 1500 rpm. Comparing the individual methods (Multi-scale 1D ResNet, LSTM-FCN, and MiniRocket), we see that the multi-scale 1D ResNet performs the best on average with the lowest standard deviation across the five random folds, followed by MiniRocket and LSTM-FCN. Despite the fact that LSTM-FCN and MiniRocket's performances were slightly lower than Multi-scale 1D ResNet, we still observed a gain in performance when we took an ensemble across these three methods, resulting in an overall accuracy to 98.86 and 98.85 for the ensembling via average operation and ensembling via the max operation respectively. Similarly, Table~\ref{tab:2700_results} shows the test scores of our experiments for the data with operating condition 2700 rpm. Upon comparing the individual methods on this dataset, we found that LSTM-FCN performed the best on average across the five random folds, followed by Multi-scale 1D ResNet and MiniRocket, while the standard deviation was best for Multi-scale 1D ResNet among all. A similar gain was achieved when taking an ensemble across these methods, with average and max ensemble resulting in an overall accuracy of 98.96 and 98.95, respectively. We also repeated the experiments in Table~\ref{tab:1500_results} and~\ref{tab:2700_results} with another separate five-folds that were split with a different random seed, and we observed similar results, with the five approaches ranking in a similar order. Based on these results, we conclude that in the context of this dataset, performing an ensemble of the three approaches via taking the average would result in the best performance in terms of accuracy. 

\subsection{Training time comparison}
We would also highlight that based on the last column in Table~\ref{tab:1500_results} and~\ref{tab:2700_results}, the Multi-scale 1D ResNet takes approximately 300 seconds longer to train in comparison to LSTM-FCN and MiniRocket. Note that both the Ensemble method has no timing as the time taken for the operation of taking the average or maximum probability is negligible, and the cost of having an ensemble model is simply the sum of the three individual models' training time. In our experiments, both deep-learning-based methods, Multi-scale 1D ResNet and LSTM-FCN, were performed on a workstation with a six-core Intel CPU and an NVIDIA TITAN Xp GPU with 12 GB memory. In contrast, the MiniRocket experiments were performed on an 8-core M2 CPU as it does not benefit from GPU acceleration. Hence, from an end-user point of view, if accuracy is highly desirable and has access to computational resources, then using an ensemble would enable a high accuracy. On the other hand, if there are constraints on computational resources, using MiniRocket would also yield a competitive classification performance.

\subsection{Additional analysis and discussion}
\begin{figure}[!h]
    \centering
    \includegraphics[width=0.59\columnwidth]{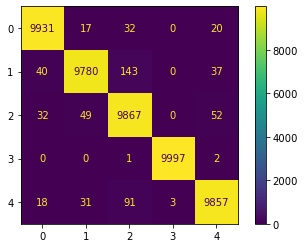}
    \caption{Confusion Matrix for Ensemble (Average) method on operating condition 1500 rpm}
    \label{fig:cm_1500}
\end{figure}

\begin{figure}[!h]
    \centering
    \includegraphics[width=0.59\columnwidth]{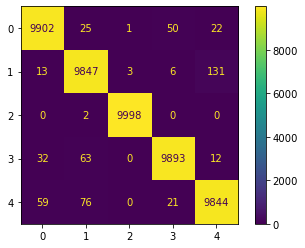}
    \caption{Confusion Matrix for Ensemble (Average) method on operating condition 2700 rpm }
    \label{fig:cm_2700}
\end{figure}

In this section, we further analyze and compare the performance of the ensembling strategy via the confusion matrices (Ground truth shown on the horizontal axis vs. predictions shown on the vertical axis) over the five-fold cross-validation. From Figure~\ref{fig:cm_1500}, we observe that for operating condition 1500 rpm, our ensemble approach almost perfectly classifies the Chipped category (Class 3) as no other class is predicted as Chipped, whereas the Crack category (Class 1) appears to be relatively the most difficult category to classify. Interestingly, for operating condition 2700 rpm, we observe that our ensemble approach almost perfectly classifies for Surface Wear category (Class 2) since no other class is predicted as Surface Wear, and performs similarly for the other classes, as seen in Figure~\ref{fig:cm_2700}. These insights could potentially be leveraged in the future to design a better loss function or data augmentation strategies to improve the method's accuracy.

\section{Conclusion}

In conclusion, our approach for fault detection using vibration signals has shown great results. We have utilized three different models, namely, Multi-scale 1D ResNet, LSTM-FCN, and MiniRocket, and each of them have performed well on their own. However, the Ensemble (Average) approach of taking the maximum of the average probability, has pushed the accuracy even further. It is worth noting that each model has its own strengths, they all use convolution kernels in different ways to extract information from the time series data, and the ensemble approach allows us to make use of the different aspects learned in each model. For future work, we plan to add more models to the ensemble, incorporate few-shot-learning techniques for applications where training data is not as easily accessible, explore functional approaches, and further leverage domain knowledge to improve the accuracy of our approach. Another line of study includes optimizing and reducing the computational complexity of such models so that they can easily be deployed on edge devices, which is often an important requirement in equipment prognostic and health management applications.

\bibliographystyle{plain}
\bibliography{ref}

\end{document}